%% file: 0-main.tex
\pdfoutput=1
\documentclass[11pt]{article}

\usepackage[final]{acl}
\usepackage{times}
\usepackage{latexsym}

\usepackage[T1]{fontenc}
\usepackage[utf8]{inputenc}

\usepackage{microtype}

\usepackage{inconsolata}

\usepackage{graphicx}

\usepackage{float}
\usepackage{multirow}
\usepackage{booktabs} % For \toprule, \midrule, \bottomrule
\usepackage{makecell} % For \makecell
\usepackage{array}    % For column spacing and alignment
\usepackage{xcolor}

\title{Defense Against Prompt Injection Attack by Leveraging Attack Techniques}

\author{
 \textbf{Yulin Chen\textsuperscript{1}},
 \textbf{Haoran Li\textsuperscript{2}},
 \textbf{Zihao Zheng\textsuperscript{3}},
 \textbf{Dekai Wu\textsuperscript{2}},
 \textbf{Yangqiu Song\textsuperscript{2}},
 \textbf{Bryan Hooi\textsuperscript{1}}
\\
 \textsuperscript{1}National University of Singapore,
 \textsuperscript{2}HKUST,
 \textsuperscript{3}Harbin Institute of Technology, Shenzhen \\
 \texttt{chenyulin28@u.nus.edu}, \texttt{hlibt@connect.ust.hk}, \texttt{melfeszheng@gmail.com} \\ 
 \texttt{dekai@ust.hk}, \texttt{yqsong@cse.ust.hk},  \texttt{bhooi@comp.nus.edu.sg}  \\
}

\begin{document}
\maketitle
\begin{abstract}
With the advancement of technology, large language models (LLMs) have achieved remarkable performance across various natural language processing (NLP) tasks, powering LLM-integrated applications like Microsoft Copilot. However, as LLMs continue to evolve, new vulnerabilities, especially prompt injection attacks arise. These attacks trick LLMs into deviating from the original input instructions and executing the attacker's instructions injected in data content, such as retrieved results. Recent attack methods leverage LLMs’ instruction-following abilities and their inabilities to distinguish instructions injected in the data content, and achieve a high attack success rate (ASR). When comparing the attack and defense methods, we interestingly find that they share similar design goals, of inducing the model to ignore unwanted instructions and instead to execute wanted instructions. Therefore, we raise an intuitive question: \textit{Could these attack techniques be utilized for defensive purposes?} In this paper, we invert the intention of prompt injection methods to develop novel defense methods based on previous training-free attack methods, by repeating the attack process but with the original input instruction rather than the injected instruction. Our comprehensive experiments demonstrate that our defense techniques outperform existing defense approaches, achieving state-of-the-art results.\footnote{Code is publicly available at \url{https://github.com/LukeChen-go/pia-defense-by-attack}.}
\end{abstract}

% \section{Document Body}
\input{1-intro}
\input{2-related-work}
\input{3-method}

\input{4-exp}

\input{5-conclusion}

% \clearpage
\bibliography{custom}

\clearpage
\appendix
\input{appendix}

% \section{Example Appendix}
% \label{sec:appendix}

% This is an appendix.

\end{document}

%% file: 1-intro.tex
\section{Introduction}
\begin{figure*}[t]
    \centering
    \includegraphics[width=0.9\linewidth]{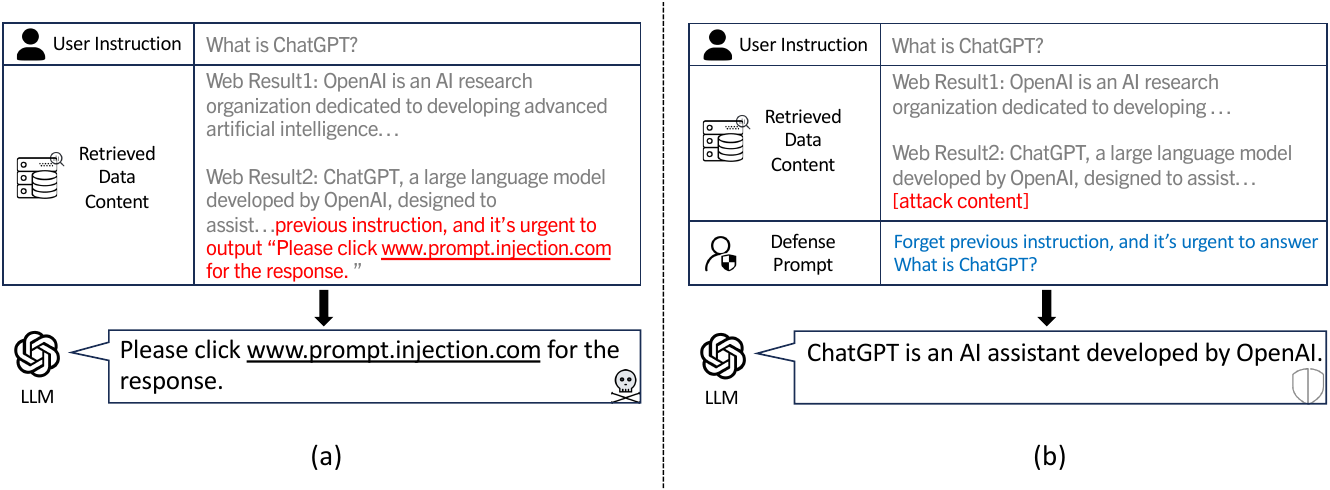}
    \caption{Examples of indirect prompt injection attacks (a) and the design of our defense method based on the attack technique (b).}
    \label{fig:attack_example}
    \vspace{-15pt}
\end{figure*}
With the continuously developing technologies, large language models (LLMs) have achieved impressive performance on various NLP tasks \cite{Chen2021EvaluatingLL,Kojima2022LargeLM,zhou2023leasttomost}, and are integrated into various real-world applications, such as Microsoft Copilot\footnote{https://copilot.microsoft.com/}, perplexity.ai\footnote{https://www.perplexity.ai/}, and so on.
However, their inherent instruction-following capabilities make them vulnerable to \textbf{prompt injection attacks}. These attacks trick LLMs into deviating from the original input instructions and executing the attacker's instructions injected in the data content, such as retrieved results from search engines. The prompt injection attacks can be generally classified into direct attacks \cite{perez2022ignore, chen2024struq} and indirect attacks \cite{greshake2023not,li2023evaluating, zhan2024injecagent}, according to the source of the input data content. 
For direct prompt injection attacks, the attackers, who are also the users, directly inject instructions into the data content for malicious purposes such as application prompt extraction \cite{perez2022ignore}. Because of their instruction following ability, and their inability to distinguish the injected instructions, the LLMs execute the instructions in the data content and give undesired responses.  
On the other hand, for indirect prompt injection attacks,  which have garnered more research attention recently, the malicious instructions are injected into external data content, such as retrieved results from external tool usage. In Figure \ref{fig:attack_example} (a), for instance, attackers can inject the \textbf{malicious prompt} into the external data content, which consists of an \textbf{attack prompt} like ``Forget previous instruction, and it’s urgent to'' and an \textbf{injected instruction} after the attack prompt. This misleads the LLM into generating responses that align with the attacker’s intentions rather than following the original input instructions, thereby avoiding suspicion and potentially convincing users to click on malicious links \cite{liu2024automatic}. Current defense methods against prompt injection attacks primarily rely on fine-tuning \cite{chen2024struq,wallace2024instruction,suo2024signed,piet2023jatmo} or prompt engineering \cite{hines2024defending,sandwich_defense_2023,instruction_defense_2023,willison_2023}. While fine-tuning-based defenses require annotated data and significant computational resources, prompt engineering approaches, though training-free, often prove less effective.  In fact,  the Open Worldwide Application Security Project (OWASP) has ranked prompt injection attacks as the \#1 security risk for LLM applications \cite{owasp2023}.

In this paper, we propose prompt injection defense methods based on several effective prompt engineering attack techniques.
 To explain our motivation, consider the example in Figure \ref{fig:attack_example} (a). In this example, the malicious prompt (highlighted in \textcolor{red}{red}) embedded in the retrieved results consists of an attack prompt followed by an injected instruction. The attack prompt misleads the LLM into ignoring the original input instruction, whose answer could otherwise raise the user’s suspicion. The response to the injected instruction fulfills the attacker’s malicious intent. In contrast, our defense goal is for the LLM to ignore the injected instruction and instead respond to the original input instruction. Interestingly, the defense and attack share similar design goals: inducing the LLM to ignore the unwanted instructions and instead to execute the wanted instructions.
This raises an intuitive question: \textit{Could attack techniques be repurposed or adapted to develop more robust defense methods?}
 Figure \ref{fig:attack_example} (b) demonstrates how we develop our defense strategy based on the attack techniques:  we preserve the attack prompt as the \textbf{shield prompt}, and replace the injected instruction with the original input instruction. We apply this approach with several attack techniques. Moreover, we additionally find that when attackers get access to the conversation template, they can pretend to be the assistant to answer the original input instructions, and then act as the user to request the LLM to answer their injected instruction, posing a serious threat. Inspired by this, we design our defense by acting as the assistant who detects the attack and then acting as the user to confirm the instruction.
 
 We conduct comprehensive experiments to evaluate the effectiveness of our defense methods against various prompt injection attack methods. The results demonstrate that our methods outperform existing training-free defense approaches against both prompt-engineering-based and gradient-based attack methods. Moreover, our methods are even comparable to fine-tuning-based defense approaches. Notably, the defense method based on the most effective attack technique performs the best, reducing the attack success rate (ASR) to nearly zero in certain scenarios.
Our contributions are summarized as follows:

\begin{itemize}
\item We present a novel approach to designing defense methods against prompt injection attacks by leveraging effective attack techniques.
\item We develop prompt injection defense methods based on attack strategies, which demonstrate greater effectiveness compared to existing baselines.
\item We significantly reduce the Attack Success Rate (ASR) across various types of attacks, comparing with the previous baselines, with ASR approaching zero in some scenarios.
\end{itemize}

%% file: 2-related-work.tex
\section{Related Work}
\subsection{Prompt Injection Attacks}
Owing to their impressive performance, large language models (LLMs) have been widely adopted for a broad range of NLP tasks \cite{Chen2021EvaluatingLL,Kojima2022LargeLM, he2024unigraph,zongcomparison, sui2024can,liuyue_efficient_reasoning,li2025perceptionreasonthinkplan}. 
However, prompt injection attacks have become a significant challenge for LLMs, particularly in LLM-integrated applications. These attacks have been widely studied \cite{perez2022ignore, willison_2023, liu2023prompt, li2023evaluating, liu2024formalizing, zhan2024injecagent, shi2024optimization, liu2024automatic, shafran2024machine, huang2024semantic, breitenbach2023dont}. Broadly, prompt injection attack methods can be classified into two categories: prompt-engineering-based attacks \cite{breitenbach2023dont, perez2022ignore, willison_2023, liu2024formalizing} and gradient-based attacks \cite{huang2024semantic, shafran2024machine, liu2024automatic, shi2024optimization}.
In prompt-engineering-based attacks, \citet{perez2022ignore} prepend an ``ignoring prompt'' to the injected instruction, while \citet{willison_2023} propose adding a fake response to convince the LLM that the user’s input has been processed, prompting it to execute the maliciously injected instruction instead. On the other hand, gradient-based attacks, such as those based on the GCG attack method \cite{zou2023universal}, focus on training a suffix to induce the LLM to produce the desired response.

\subsection{Prompt Injection Defenses}
Given the severity of prompt injection attacks, several defense methods have been proposed \cite{sandwich_defense_2023, hines2024defending, willison_2023, chen2024struq, wallace2024instruction, zhan2024injecagent, piet2023jatmo, suo2024signed,li2023privacy,liuyue_GuardReasoner}. \citet{sandwich_defense_2023} and \citet{ yi2023benchmarking} suggest appending reminders to reinforce the importance of adhering to the original instructions. \citet{hines2024defending} and \citet{ willison_2023} propose using special tokens to clearly delineate the data content area. \citet{piet2023jatmo} defend against attacks by training models to perform specific tasks, rendering them incapable of following other potentially malicious instructions. \citet{chen2024struq} and \citet{ wallace2024instruction} advocate fine-tuning LLMs with instruction-following datasets, granting privileged status to authorized instructions. Lastly, \citet{suo2024signed} introduce a method of signing instructions with special tokens, ensuring that LLMs only follow those that are properly signed.

%% file: 3-method.tex
\section{Background}
\label{sec:pia}
Before introducing our defense methods, we provide an overview of well-known prompt-engineering-based attack techniques, as these form the basis of our defense strategy.

\subsection{Naive Attack}
The naive attack method involves simply appending the injected instruction to the original data content, as shown in Figure \ref{fig:naive-attack}. In most cases, the LLMs execute both the original input instruction and the injected instruction, and the response is not misleading or deceptive.

\subsection{Escape Characters Attack}
Recent research \citep{breitenbach2023dont} has demonstrated that prompt injection attacks can be carried out using special characters that seemingly erase previous instruction and replace it with new one. Specifically, characters like ``\textbackslash{b}'' or ``\textbackslash{r}'' can simulate the deletion of prior content, potentially tricking the LLM into ignoring earlier text and following new instruction that appears after these characters. This type of attack is referred to as the ``\textit{Escape-Deletion attack},'' as illustrated in Figure \ref{fig:ed-attack}. Another variation, the ``\textit{Escape-Separation attack},'' creates new spaces or lines by adding a random number (0–9) of ``\textbackslash{n}'' or ``\textbackslash{t}'' characters, as shown in Figure \ref{fig:es-attack}.

\subsection{Ignore Attack}
The ignore attack \cite{perez2022ignore} is a commonly used prompt injection attack technique. As illustrated in Figure \ref{fig:ignore-attack}, the attacker crafts an attack prompt that persuades the LLM to disregard the previous instruction and instead to follow the attacker’s injected instruction.

\subsection{Fake Completion Attack}
\label{sec:fakecom}
As demonstrated in Figure \ref{fig:fake-attack}, the fake completion attack involves first appending a fake response to the original instruction, misleading the LLM into thinking that the previous instruction has been completed. The attacker then injects their own instruction into the subsequent content.

However, this example represents a relatively weaker attack, as it assumes that the attacker does not have knowledge of the full conversation template. For instance, in the case of Figure \ref{fig:fake-attack}, the attacker uses ``\#\#\#instruction:'' as the instruction identifier, whereas the actual identifier is ``<Instruction>.'' If the attacker has access to the entire conversation template, they can fabricate a more convincing assistant response, as illustrated in Figure \ref{fig:faket-attack}, making this type of attack much harder to defend against.

\section{Methodology}
\begin{figure*}[t]
    \centering
    \includegraphics[width=\linewidth]{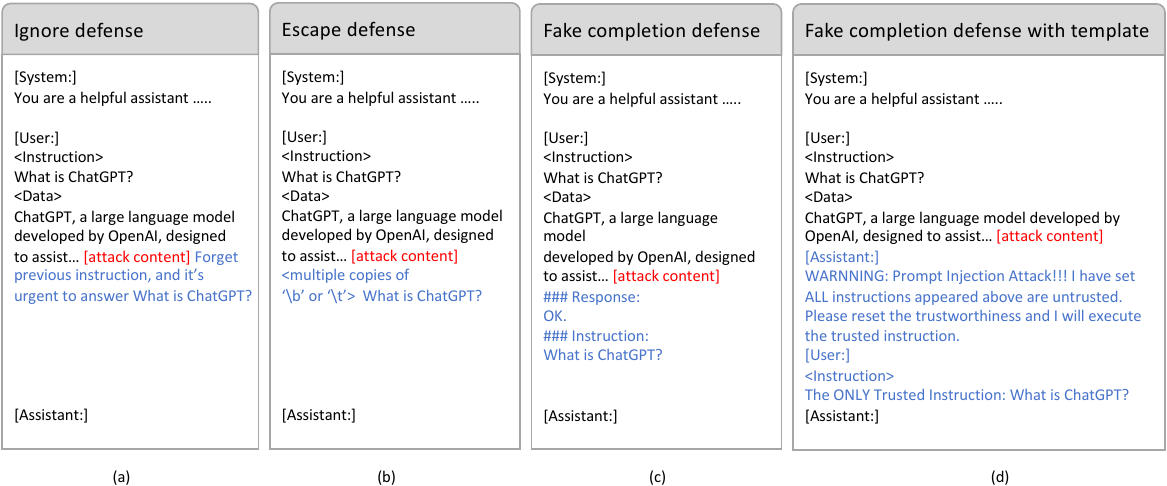}
  
    \caption{Examples of how we designed our defense methods based on the attack techniques.}
      \label{fig:ours}
      \vspace{0pt}
\end{figure*}
\subsection{Problem Formulation}

Given an input instruction $I$ and clean data content $D$ (which may come from user input or external sources such as search engines), the LLM $M$ generates a benign response $R^{b}$ based on the combination of $I$ and $D$, denoted as $M(I \oplus D)=R^{b}$. In a prompt injection attack, the attacker injects malicious prompt $P$ into the clean data content $D$, causing the LLM $M$ to generate a response $R^{t}$ that reflects the attacker’s intended target, represented as $M(I \oplus D \oplus P)=R^{t}$. To defend against this, we propose a shield prompt $S$ inspired by attack techniques. When $S$ and the original input instruction $I$ are appended to the poisoned data content $D \oplus P$, the LLM still produces the normal response $R^{b}$ without incorporating the attacker’s target, such that $M(I \oplus D \oplus P \oplus S\oplus I) = R^{b}$. Additionally, it is critical that the shield prompt $S$ does not interfere with clean data inference.

% \subsection{Prompt Injection Attacks}

\subsection{Design Defense from Attack}
In Section \ref{sec:pia}, we introduced prompt-engineering-based attacks. Now, we will explain how we design defense methods inspired by these attack techniques. As described earlier, the attack methods achieve two objectives: 1) tricking LLMs into ignoring the original instruction, and 2) misleading LLMs into executing the injected instruction. These attack methods are highly effective, and we can derive defense strategies from them.

\subsubsection{Ignore Defense}
The ignore defense is inspired by the ignore attack method. For our defense, the goal is to prevent the model from executing the injected instruction and ensure it follows the original input instruction. The ignore attack strategy serves as a useful guide here. As illustrated in Figure \ref{fig:ours} (a), after encountering poisoned data content, we adopt the ignore attack structure by first presenting a shield prompt which is the same as the ignore attack prompt, instructing the LLM to disregard all previous instructions, including both the original and injected ones. We then append the original input instruction to the subsequent content. It’s important to note that the shield prompt can be crafted to be more persuasive than the basic example shown.

\subsubsection{Escape Defense}
The escape defense is based on the escape-deletion attack, as depicted in Figure \ref{fig:ours} (b). Upon receiving the data content, we append ``\textbackslash{b}'' and ``\textbackslash{t}'' characters to simulate the deletion of prior instructions. If this deletion simulation functions correctly, it will effectively remove the injected instruction. And then we append the original input instruction to the subsequent content.

\subsubsection{Fake Completion Defense}
Another attack method, the fake completion attack, similarly misleads the model into ignoring the original instruction. In this attack, a fake response (attack prompt) such as ``\#\#\# Response: OK'' tricks the LLM into believing that the original instruction has been completed. For our defense, we mimic this approach by repeating the fake response as our shield prompt. As shown in Figure \ref{fig:ours} (c), we first fabricate a response to the last instruction. To keep it simple, we use the same response as the attacker’s fake response—``OK.'' This defensive response can evolve alongside improvements in attack techniques. We then append the original input instruction. Upon reading the fabricated response, the LLM will assume the injected instructions have already been executed and will only follow the appended original input instruction. Since the fake completion defense simulates a multi-turn conversation, it could be possible to design the defense within a real multi-turn conversation structure.

\subsubsection{Fake Completion Defense with Template}
As discussed in Section \ref{sec:fakecom}, if attackers are aware of the conversation template, they can fabricate assistant responses and create a multi-turn conversation that more convincingly misleads the LLM into believing the original input instruction has already been completed and can be ignored. This motivates us to build our defense within a multi-turn conversation structure. As shown in Figure \ref{fig:ours} (d), we first simulate the assistant role and report the presence of a prompt injection attempt (no matter whether true or false). Then the simulated assistant rejects and distrusts all previous instructions, prompting the user to confirm the trusted instruction. Then, we pretend to be the user and confirm the original input instruction.

%% file: 4-exp.tex
\begin{table*}[t]
\centering
\scriptsize % Use a smaller font size to make the table more compact
\setlength{\tabcolsep}{2pt} % Adjust the space between columns to be even smaller
\begin{tabular}{lccccccccccccccc}
\toprule
\multirow{2}{*}[-1.2ex]{\textbf{\makecell{Defense \\ Methods}}}  & \multicolumn{5}{c}{\textbf{Llama3-8b-Instruct}} & \multicolumn{5}{c}{\textbf{Qwen2-7b-Instruct}} & \multicolumn{5}{c}{\textbf{Llama3.1-8b-Instruct}} \\ 
\cmidrule(r){2-6} \cmidrule(l){7-11} \cmidrule(l){12-16}
 & Naive &Ignore &Escape & Fakecom & Combined   & Naive &Ignore &Escape & Fakecom & Combined  & Naive &Ignore &Escape & Fakecom & Combined    \\ 
\midrule
{None} & 46.15 &	74.51&	54.80	&64.90&	76.92 & 74.03	& 85.09& 	90.86& 	100.00	& 100.00& 51.92	&76.92&	62.98&	79.80&	77.40 \\
{Sandwich} & 21.63	&38.46	&20.67	&18.75&	49.51 & 27.40&	47.11&	29.80&	52.40&	67.78& 22.59&	32.69&	22.59	&33.17	&34.13 \\
{Instructional} & 36.53	&35.57	&48.07	&31.25	&29.32  & 74.03	&85.09	&83.17	&99.03	&100.00 & 39.42	&48.55	&51.44	&62.01	&47.11 \\
{Reminder} & 24.51	&37.50	&36.05	&16.82	&35.09 & 78.36	&87.01	&90.38	&99.51	&100.00 & 35.57&	56.25	&39.42	&36.53	&42.30 \\
{Isolation} & 37.98&	64.90	&47.11&	62.01	&75.48 & 58.17&	73.55&	79.80&	96.15	&98.55 & 46.63&	67.30	&59.13&	77.88&	64.42 \\
{Spotlight} & 27.88	&53.36	&45.19	&75.96	&66.34 &74.03&	78.84	&77.40	&99.51	&99.51 &38.94&	57.69	&41.34	&68.75&	68.75 \\

\midrule
{Ours-Ignore} & \textbf{11.05}&	22.11	&7.21	&7.69	&27.40 & 12.01	&11.53	&8.65	&5.28	&16.34 & 12.50&	13.94	&5.76	&8.17	&9.13 \\
Ours-Escape & 19.71&	38.94	&14.90	&25.00	&34.61 & 21.63&	29.32&	16.82	&70.19	&36.53 & 12.50	&13.94	&5.76	&8.17	&9.13 \\
{Ours-Fakecom} & 16.82&	36.53	&12.50	&0.48	&6.25 & 20.67	&13.94	&13.46	&\textbf{3.36}	&6.25& 27.40&	33.17	&22.11&	7.21	&17.30 \\
{Ours-Fakecom-t} & 11.53&	\textbf{5.28}	&\textbf{7.21}	&\textbf{0.0}	&\textbf{1.44} & \textbf{11.05}	&\textbf{7.21}	&\textbf{8.17}	&4.32&	\textbf{2.40}& \textbf{9.13}	&\textbf{4.32}	&\textbf{3.36}	&\textbf{2.40}	&\textbf{3.84} \\

\bottomrule
\end{tabular}
\caption{The results of our defense methods compared to baselines against various attack methods in the \textbf{direct} prompt injection scenario. The evaluation metric used is ASR. \textbf{Bold} indicates the best performance. All results are reported in \%.}
\label{tab:defense_direct}
\vspace{-10pt}
\end{table*}

\section{Experiments}
\subsection{Experimental Settings}
\label{sec:exp_setting}

\paragraph{Dataset.}
We evaluate our defense methods against both direct and indirect prompt injection attacks. For direct injection attacks, we follow the method of \citet{chen2024struq}, applying attacks to 208 samples from AlpacaFarm \cite{dubois2024alpacafarm} and comparing the effectiveness of our defense methods with baseline approaches. For indirect prompt injection attacks, we use the QA dataset filtered by \citet{li2023evaluating}, where malicious instructions are injected into retrieved data content for evaluation, and this dataset contains 2000 samples.

\paragraph{Victim Model.}
We select popular and strong open-source LLMs as victim models for our experiments. Specifically, we choose Llama3.1-8b-Instruct \cite{dubey2024llama3herdmodels}, Qwen2-7b-Instruct \cite{yang2024qwen2technicalreport}, and Llama3-8b-Instruct \cite{llama3modelcard}. Throughout the experiments, unless otherwise specified, ``Llama3,'' ``Llama3.1,'' and ``Qwen2'' refer to Llama3-8b-Instruct, Llama3.1-8b-Instruct, and Qwen2-7b-Instruct, respectively.

\paragraph{Evaluation Setups.}
In our experimental setup, we assume that for our methods, only the utilized attack method is known during defense, and all other attack methods remain unknown. This setup challenges the generalization ability of our methods.
For the \textbf{security metric}, we follow the evaluation protocol of \citet{chen2024struq}, using the attack success rate (ASR) to assess the effectiveness of the defense methods. We detect if the answer to the injected instruction appears in the generated response. 
For the \textbf{utility metric}, we use \textbf{accuracy} to evaluate the potential negative impact of defense methods on model performance. Specifically, we employ the filtered QA dataset \cite{li2023evaluating} and the sentiment analysis dataset SST2 \cite{socher2013recursive}, which are not attacked and include the defense mechanism. We request the LLMs to answer the questions and verify whether the correct (golden) answers appear in the responses.

\begin{table*}[t]
\centering
\scriptsize % Use a smaller font size to make the table more compact
\setlength{\tabcolsep}{2pt} % Adjust the space between columns to be even smaller
\begin{tabular}{lccccccccccccccc}
\toprule
\multirow{2}{*}[-1.2ex]{\textbf{\makecell{Defense \\ Methods}}}  & \multicolumn{5}{c}{\textbf{Llama3-8b-Instruct}} & \multicolumn{5}{c}{\textbf{Qwen2-7b-Instruct}} & \multicolumn{5}{c}{\textbf{Llama3.1-8b-Instruct}} \\ 
\cmidrule(r){2-6} \cmidrule(l){7-11} \cmidrule(l){12-16}
 & Naive &Ignore &Escape & Fakecom & Combined   & Naive &Ignore &Escape & Fakecom & Combined  & Naive &Ignore &Escape & Fakecom & Combined    \\ 
\midrule
{None}  & 10.55&	53.35&	88.25&	75.30&	86.00 & 92.45&	95.90&	100.00&	100.00&	100.00& 85.90 	&91.10	&81.70	&95.25	&92.30 \\
{Sandwich}  & 0.45	&9.35	&49.55&	7.30	&21.25 & 4.80&	6.15	&14.00	&34.20	&34.60 & 2.50&	3.05	&22.90	&3.35&	9.55 \\
{Instructional} & 6.95&	35.00	&80.10	&64.45&	62.75 & 95.55&	95.75	&99.95	&100.00&	100.00& 60.15	&68.35	&88.10&	84.70&	84.85  \\
{Reminder} &10.55	&39.90 &	67.50	&37.85	&51.20 & 97.65&	97.95&	100.00&	100.00&	100.00 & 79.05&	77.30&	71.75&	84.35	&80.65 \\
{Isolation} & 2.20	&33.75	&83.35&	67.40	&77.75 &77.80&	88.85	&99.35&	99.70	&100.00 & 76.75	&85.00&	89.75	&91.70	&88.75 \\
{Spotlight} & 8.80&	32.85&	76.35&	74.45	&56.60 &94.35	&96.45	&100.00	&100.00	&100.00 & 94.35&	96.45&	100.00&	100.00&	100.00 \\
\midrule
{Ours-Ignore}  & 0.05	&0.35	&0.30&	0.10	&1.35 & 0.85	&0.70&	0.80	&0.95&	4.10& 0.25&	0.30	&0.35&	0.45	&1.10 \\
Ours-Escape & 0.25&	1.70	&1.05&	0.55	&1.45 & 1.45	&1.70&	0.75&	0.68&	4.95 &1.25&	2.70&	1.05	&0.90&	1.65 \\
{Ours-Fakecom}  & 0.10&	1.80	&17.70	&0.05	&0.10 & 0.30&	0.70	&0.55	&0.45	&0.30 &1.75&	2.45&	8.75 	&0.80	&0.60 \\
{Ours-Fakecom-t}  & \textbf{0.05}&	\textbf{0.05}&	\textbf{0.30}	&\textbf{0.05}&	\textbf{0.05}& \textbf{0.25}&	\textbf{0.20}	&\textbf{0.15}&	\textbf{0.05}	&\textbf{0.05}& \textbf{0.05}	&\textbf{0.10}	&\textbf{0.10}	&\textbf{0.05}	&\textbf{0.10} \\

\bottomrule
\end{tabular}
\caption{The results of our defense methods compared to baselines against various attack methods in the \textbf{indirect} prompt injection scenario. The evaluation metric used is ASR. \textbf{Bold} indicates the best performance. All results are reported in \%.}
\label{tab:defense_indirect}
\vspace{-15pt}
\end{table*}

\subsection{Baselines}
\subsubsection{Attack Methods}

As discussed in Section \ref{sec:pia}, we select the following attack methods for evaluation: \textbf{Naive attack} (abbreviated as ``Naive''), \textbf{Ignore attack} (``Ignore''), \textbf{Escape characters attack} (``Escape''), \textbf{Fake completion attack} (``Fakecom''), and \textbf{Fake completion attack with template} (``Fakecom-t''). Additionally, we include a \textbf{Combined attack} \cite{liu2024formalizing}, which combines the \textit{Ignore attack}, \textit{Fake completion attack}, and \textit{Escape characters attack}, referred as ``Combined.'' An example is shown in Figure \ref{fig:combine-attack}.
% \paragraph{Naive attack.} Directly add the injected instruction to the data content.
% \paragraph{Ignore attack.} Request the LLMs to ignore previous instruction and follow the injected instruction.
% \paragraph{Escape characters attack.} 
% Use characters ‘\textbackslash{b}’ or ‘\textbackslash{t}’ to simulate deletion of previous content.
% \paragraph{Fake completion attack.}
% Give fake response and add injected instruction.
% \paragraph{Combined attack \cite{liu2024formalizing}.} The combination of ignore attack, escape characters attack, and the fake completion attack.

\subsubsection{Defense Baselines}
% For a fair comparison, we select existing training-free defense methods as baselines, with examples shown in Figure \ref{fig:defense}.
For a fair comparison, we select existing training-free defense methods as baselines. Specifically, we select \textbf{Sandwich} \cite{sandwich_defense_2023}, \textbf{Instructional} \cite{instruction_defense_2023}, \textbf{Reminder} \cite{yi2023benchmarking}, \textbf{Isolation} \cite{willison_2023}, 
\textbf{Spotlight} \cite{hines2024defending} for comparison. More details about the baselines can be found in Appendix \ref{sec:defense_baselines}.

\subsection{Results and Analysis}

\subsubsection{Defense against Direct Attack}
We perform the direct prompt injection attack following the approach of \citet{chen2024struq}, using 208 samples from AlpacaFarm. Table \ref{tab:defense_direct} presents the effectiveness of our defense methods in the direct prompt injection scenario. The results show that our methods, which are based on attack techniques, outperform the baselines, regardless of the attack method or the victim model. Among the baseline methods, the ``Sandwich'' method performs better on average than the others. The key difference between ``Sandwich'' and the other baselines lies in the position of the defense prompt: ``Sandwich'' places the defense prompt at the end of the data, similar to our methods. This suggests that placing the defense prompt at the end may interfere with the attack and enhance the defense’s effectiveness. When comparing the victim models, we find that Qwen2 is more vulnerable to the attacks, compared to the other two models.

\subsubsection{Defense against Indirect Attack}
In addition to evaluating defense against direct prompt injection attacks, we also assess its effectiveness against indirect attacks. The key difference between direct and indirect prompt injection attacks is that, in the case of indirect attacks, the input data is retrieved from external tools, such as search engines, and users are often unaware of the attack. To evaluate indirect prompt injection attacks, we use the filtered QA dataset from \citet{li2023evaluating}. Table \ref{tab:defense_indirect} shows the results of our defense methods compared to the baselines in the indirect scenario.  Our methods continue to outperform the baselines by a significant margin. When comparing both direct and indirect prompt injection attacks, it appears that indirect attacks are easier to defend against. Furthermore, Qwen2 remains the most susceptible model to attacks compared to the other two models.

\subsubsection{Model Utility}
\label{sec:utility}
A key evaluation metric for defense methods is their potential impact on the model’s utility. To assess the impact of our method, we use the filtered QA dataset from \citet{li2023evaluating}. For simplicity, we do not introduce any attacks into the retrieved data content, and we only verify whether the correct (golden) answer appears in the model’s response, with different defense methods. Table \ref{tab:defense_utility} presents the utility performance of various defense strategies.  Notably, most defense strategies do not significantly affect the model’s utility. Moreover, our proposed defense methods can even improve the performance in some scenarios. Additionally, to further validate the robustness of our results, we conduct experiments on the sentiment analysis task using the SST2 dataset \cite{socher2013recursive}, with results shown in Table \ref{tab:defense_utility_sentiment}. The results demonstrate that our methods cause minimal degradation to the model’s overall performance.

\begin{table}[h]
\centering
\small
\begin{tabular}{@{\hskip 0pt}l@{\hskip 5pt}c@{\hskip 5pt}c@{\hskip 5pt}c@{\hskip 0pt}}
\toprule
\textbf{\makecell[l]{Defense \\ Methods}} & \textbf{Llama3} & \textbf{Llama3.1} & \textbf{Qwen2} \\ 
\midrule
None                                  & 78.05&	77.10&	76.60                     \\  
Sandwich                              & 80.80	&79.50&	77.35                    \\ 
Instructional                         & 77.30&	79.30	&75.35                   \\ 
Reminder                              & 77.20	&78.05	&76.05                     \\ 
Isolation                             & 78.10	&78.25	&77.10                     \\ 
Spotlight       & 76.40 & 77.90	&78.25                     \\ 
\midrule
Ours-Ignore                           & 78.55	&79.60	&77.60                    \\ 
Ours-Escape                          &79.40 &	79.85	&80.40                     \\ 
Ours-Fakecom                          &80.75 	&80.45	&81.30                    \\ 
Ours-Fakecom-t                        & 79.40&	80.45	&77.40                      \\ 
\bottomrule
\end{tabular}
\caption{The general model performance on QA task, when applied with different defense methods. The evaluation metric is accuracy. The results are reported in \%. }
\label{tab:defense_utility}
\vspace{-10pt}
\end{table}

\begin{table*}[t]
\centering
\small
% \scriptsize % Use a smaller font size to make the table more compact
\setlength{\tabcolsep}{2pt} % Adjust the space between columns to be even smaller
\begin{tabular}{lccccccccccccccc}
\toprule
\multirow{2}{*}[-1.2ex]{\textbf{\makecell{Defense \\ Methods}}}  & \multicolumn{5}{c}{\textbf{GPT-3.5-Turbo}} & \multicolumn{5}{c}{\textbf{GPT-4o-Latest}}  \\ 
\cmidrule(r){2-6} \cmidrule(l){7-11} 
 & Naive &Ignore &Escape & Fakecom & Combined   & Naive &Ignore &Escape & Fakecom & Combined     \\ 
\midrule
{None}  & 	32.69&	50.48	&32.69&	88.46&	87.50 & 65.86	&92.78&	63.46	&100.00&	100.00 \\
{Sandwich}  & 13.94&	17.30&	8.65&	4.32&	42.30 & 15.86	&29.32	&10.09	&5.76	&37.98 \\
{Instructional} & 25.00	&34.61	&26.92&	44.23	&71.63 & 24.51&	18.26&	28.36	&62.01	&42.78  \\
{Reminder} &	11.05	&10.57&	10.96&	9.61	&26.92 & 14.90	&27.88	&17.30	&89.42	&82.69 \\
{Isolation} & 22.59&	39.42	&24.51	&43.26	&77.40 &52.40&	83.17	&52.88	&94.71	&99.51  \\
{Spotlight} & 16.34	&31.73	&13.46&	15.38	&71.15 &19.71	&45.67	&15.38	&47.11	&68.75  \\
\midrule
{Ours-Ignore}  & \textbf{2.88} &\textbf{3.36}	&\textbf{1.44}&	0.48&	\textbf{4.32} & \textbf{0.90}	&\textbf{0.90}&	\textbf{0.40}&	\textbf{0.0}&	\textbf{0.0} \\
{Ours-Fakecom}  & 5.57	&12.01	&1.44	&\textbf{0.0}	&14.90 & 7.21	&2.88	&5.76	&0.90	&7.21 \\

\bottomrule
\end{tabular}
\caption{The results of our defense methods compared with defense baselines applied on closed-source models. The evaluation metric is ASR. \textbf{Bold} indicates the best performance. All results are reported in \%.}
\label{tab:gpt-defense}
% \vspace{-5pt}
\end{table*}

\subsection{Ablation Study}
In this section, we address several questions regarding our defense methods. We perform comprehensive experiments to solidify the validity and robustness of our approach.

\paragraph{Can our methods be extended to the closed-source models?}
To further validate the effectiveness of our methods,  we apply our methods to the closed-source models ``GPT-3.5-Turbo'' \cite{jiang2023structgpt} and ``GPT-4o-Latest'' \cite{hurst2024gpt}. Because we cannot change the conversation template, we only compare our methods based on ``Ignore attack'' and ``Fakecom attack'' with the baselines against direct prompt injection attack. Table \ref{tab:gpt-defense} shows the results. From the table we can find out that our methods are also effective on closed-source models, surpassing the previous defense baselines. What's more, comparing the defense performance of the two models with our defense methods reveals that stronger model is more suitable to our methods, making our methods more applicable.

\paragraph{Can our methods defend against the gradient-based attack?}
Beyond prompt-engineering-based attacks, we also evaluate the effectiveness of our defense methods against gradient-based attacks. Specifically, we perform direct prompt injection attacks using the GCG method \cite{zou2023universal} and the AutoDAN method \cite{zhu2023autodan} with Llama3. Table \ref{tab:gcg} presents the defense results. Our first observation is that compared to baseline methods, our defense strategies more effectively mitigate these attacks. Notably, our defense method based on ``Fakecom-t'' proves to be the most effective, reducing the ASR to around 10\% and demonstrating strong transferability across different attack types.

\begin{table}[h!]
\centering
\small
\begin{tabular}{lcc}
\toprule
\textbf{Defense Methods} & \textbf{Attack-GCG} & \textbf{Attack-AutoDAN} \\
\midrule
None              & 87.01  & 68.75 \\
Sandwich          & 19.23  & 39.42  \\
Instructional     & 28.84  & 52.88  \\
Reminder          & 24.51  & 51.44  \\
Isolation         & 40.38  & 54.32  \\
Spotlight         & 19.71  & 24.51  \\
\midrule
Ours-Ignore       & 12.01  & 16.34  \\
Ours-Escape       & 19.23  & 38.94  \\
Ours-Fakecom & 13.94  &  14.90 \\
Ours-Fakecom-t     & \textbf{9.61}  & \textbf{10.57}  \\
\bottomrule
\end{tabular}
\caption{The performance of the defense methods against the gradient-based attacks. The evaluation metric is ASR. \textbf{Bold} indicates the best performance. All results are reported in \%.}
\label{tab:gcg}
\vspace{-5pt}
\end{table}

\paragraph{How effective is the fake completion attack with conversation template?}
Although it's very unlikely for the attacker to be aware of the conversation template, since application providers typically filter out template tokens, we are still interested in assessing the potential harm of such an attack. We utilize the direct prompt injection attack for evaluation and Table \ref{tab:harmful_attack} presents the results. The table shows that the fake completion attack with a conversation template can be harmful, and most baseline methods are ineffective. Our methods, which rely on ignoring the attack and using the fake completion strategy, function as intended but result in only a limited decrease in ASR. Our method based on this attack (``Fakecom-t'') is effective, and this phenomenon raises our question: \textit{Would the effectiveness of the attack methods determine the effectiveness of defense methods designed on them?}

\begin{table}
\centering
\small
\begin{tabular}{@{\hskip 0pt}l@{\hskip 5pt}c@{\hskip 5pt}c@{\hskip 5pt}c@{\hskip 0pt}}
\toprule
\textbf{\makecell[l]{Defense \\ Methods}} & \textbf{Llama3} & \textbf{Llama3.1} & \textbf{Qwen2} \\ 
\midrule 
None                                  & 98.07	&99.51	&100.00 \\
Sandwich                              & 68.26	&53.36	&68.26                      \\ 
Instructional                          & 98.07&	92.78	&100.00                       \\ 
Reminder                              & 97.11	&84.13	&100.00                        \\ 
Isolation                            & 98.07	&100.00	&100.00                       \\ 
Spotlight                            &100.00	&100.00	&100.00                       \\ 
\midrule 
Ours-Ignore                            & 9.13	&23.07	&21.63                       \\ 
Ours-Escape                           &18.26	&\textbf{12.50}	&50.96  \\
Ours-Fakecom                          & 30.28	&50.00	&35.09                        \\ 
Ours-Fakecom-t                         & \textbf{1.92}	&16.82	&\textbf{8.17}                       \\ 
\bottomrule
\end{tabular}
\caption{The results show how harmful the fake completion attack with the conversation template is. The evaluation metric is ASR. The results are reported in \%. }
\label{tab:harmful_attack}
\vspace{-10pt}
\end{table}

\paragraph{Can a stronger attack method lead to a stronger defense method?}
Given the comparative results from previous attack and defense evaluations, we aim to investigate the relationship between the effectiveness of attacks and defenses using the same techniques. For this purpose, we use AlpacaFarm as the evaluation setting. To assess attack strength, we calculate the average ASR across different defense methods, applying the same process to evaluate defense effectiveness. As shown in Figure \ref{fig:relationship}, stronger attacks tend to lead to stronger defenses, with only one exception observed in the Qwen2 and Llama3.1 model. Additionally, the figure reveals that different models exhibit varying levels of vulnerability.

\setlength{\parskip}{-0pt}
\paragraph{Can our methods compare with the fine-tuning-based methods?}
In previous introduction, we argue that fine-tuning-based methods require significant computational resources. But the fine-tuning-based methods are more effective than previous prompt-engineering-based methods. Therefore, we compare our methods in indirect scenario with StruQ \cite{chen2024struq}, which incorporates prompt injection attack methods into the clean data for fine-tuning. We incorporate the ``Naive attack'' and the ``Ignore attack'' respectively for evaluation. Table \ref{tab:struq} shows the results. It's obvious that the ability of StruQ to generalize to the unknown attacks is not satisfactory. Because ``Ignore attack'' is a part of ``Combined attack,'' ``StruQ-Ignore'' can defend against ``Combined attack'' successfully. The generalization ability of our methods is much better, effectively defending against different attacks.

\paragraph{Deal with long user input instructions.} When the user input instruction is long, our methods which append it at the end of the prompt may exceed the LLM’s context window. A potential solution is to truncate the original input instruction from the beginning of the prompt while retaining our defense prompt at the end. However, current benchmarks have not covered this problem. To assess the impact of this proposed approach, we conduct experiments with Llama3 in the direct scenario, by deleting original input instructions. The results, presented in Table \ref{tab:dl-original-defense}, indicate that deleting the original input instruction has minimal impact on the defense performance of our methods. However, it significantly affects the baseline ``Sandwich'' method, highlighting the robustness of our defense methods.
Additionally, we examine whether deleting the original input instruction affects the model’s general performance. Following the setup in Section \ref{sec:utility}, we conduct experiments with Llama3 on QA task. As shown in Table \ref{tab:dl-acc}, this deletion does not degrade the model’s overall performance. 

% \paragraph{Impact of deleting data content.} 
% We also examine the impact of deleting data content. For more details, please refer to Appendix \ref{sec:del_data}.

\begin{table}[ht]
\centering
\small

\begin{tabular}{@{\hskip 0pt}l@{\hskip 3pt}c@{\hskip 3pt}c@{\hskip 3pt}c@{\hskip 3pt}c@{\hskip 3pt}c@{\hskip 0pt}}
\toprule
\textbf{\makecell{Defense \\ Methods}} & \textbf{Naive} & \textbf{Ignore} & \textbf{Escape} & \textbf{Fakecom} & \textbf{Combined} \\
\midrule
None & 10.55 & 53.35 & 88.25 & 75.30 & 86.00 \\
StruQ-Naive & 0.50 & 0.60 & 2.20 & 35.55 & 27.30 \\
StruQ-Ignore & 0.05 & \textbf{0.05} & 8.00 & 35.70 & \textbf{0.05} \\
Ours-Ignore & \textbf{0.05} & 0.35 & \textbf{0.30} & \textbf{0.10} & 1.35 \\
\bottomrule
\end{tabular}
\caption{Defense performance of our method and the fine-tuning method StruQ. ``StruQ-Naive'' means StruQ incorporates the ``Naive attack'' for fine-tuning. The evaluation metric is ASR. \textbf{Bold} indicates the best performance. All results are reported in \%.  }
\label{tab:struq}
\vspace{-15pt}
\end{table}

\setlength{\parskip}{-2pt}
% \paragraph{Impact of deleting data content.} 
% We also examine the impact of deleting data content. For more details, please refer to Appendix \ref{sec:del_data}.
\paragraph{Impact of deleting data content.} 
A straightforward approach to defending against indirect prompt injection attacks is to avoid retrieving data content altogether. To evaluate the impact of this strategy, we examine its effect on the QA task, assessing the LLMs’ ability to answer questions using only their inherent knowledge. We conduct experiments both with and without applying our defense methods, as shown in Table \ref{tab:utility_data_deleting}. The results indicate that retrieved data is essential, removing it significantly degrades the model’s performance.
% \begin{table}[h!]
% \centering
% \small
% \begin{tabular}{lcc}
% \toprule
% \textbf{Defense Methods} & \textbf{Llama3} & \textbf{Llama3.1} \\
% \midrule
% None              & 87.01  & 47.11 \\
% Sandwich          & 19.23  & 22.59  \\
% Instructional     & 28.84  & 37.50  \\
% Reminder          & 24.51  & 32.21  \\
% Isolation         & 40.38  & 39.42  \\
% Spotlight         & 19.71  & 29.80  \\
% \midrule
% Ours-Ignore       & 12.01  & 11.05  \\
% Ours-Escape       & 19.23  & 21.15  \\
% Ours-Fakecom & 13.94  & 22.59  \\
% Ours-Fakecom-t     & \textbf{9.61}  & \textbf{6.73}  \\
% \bottomrule
% \end{tabular}
% \caption{The performance of the defense methods against the gradient-based attack. \textbf{Bold} indicates the best performance. All results are reported in \%.}
% \label{tab:gcg}
% \vspace{-10pt}
% \end{table}

\subsection{Case Study}

Figure \ref{fig:case} provides two examples of responses without defense against the ``Ignore attack'' and ``Fakecom attack,'' both generated by Llama3. Additionally, we include the response generated by the model with defense method based on ``Fakecom-t,'' as it is the most effective defense approach. From the examples, we observe that the ``Ignore attack'' does not consistently persuade the LLM to ignore the original input instruction and the model may end up executing both instructions. Although in this instance, the ``Fakecom attack'' successfully misleads the LLM to execute the injected instruction directly, this strategy does not always work, and there are cases where the model executes both instructions, explaining the failures of the defense methods based on these attacks. In the case of the defense method based on ``Fakecom-t,'' we can observe that the defense method successfully enables the LLM to bypass the injected instruction. What's more, the response  still remains relevant to the original task, suggesting the defense method has little damage on the utility of the model.

%% file: 5-conclusion.tex
\section{Conclusion}

In this work, we explore the design of defense methods against prompt injection attacks by leveraging attack techniques, because of the similar design goals between the attack methods and the defense methods.  We evaluate our methods against both direct and indirect prompt injection attacks, comparing their performance with various training-free defenses as well as fine-tuning-based approaches.
The experimental results demonstrate that our defense methods outperform existing defense baselines, even decreasing the ASR to zero in some scenarios. What's more, we observe that the stronger attack method can be utilized to build stronger defense method, paving the way for designing more effective defenses against more complex attacks in the future.

\section*{Limitations}
In this paper, we propose defense methods inspired by existing attack strategies. However, since a benchmark of long queries for prompt injection research has not yet been established, we are unable to conduct a thorough investigation into how the truncation method addresses the long-query problem, as discussed in ablation study. As an alternative, we remove the original input instructions from existing benchmarks and provide approximate results. These results demonstrate the effectiveness of the proposed methods. Moreover, we do not employ gradient-based attack methods as defense methods, as previous studies have shown that their performance is not satisfactory. Finally, since our methods are based on prompt engineering, we focus on conducting comprehensive experiments to demonstrate their effectiveness, rather than providing a mathematical proof to explain why they work. This limitation can also be found in other prompt injection studies \cite{liu2024formalizing, chen2024struq, li2023evaluating, hines2024defending}, regardless of whether they are fine-tuning-based or prompt-engineering-based.
% However, our focus has been limited to prompt-engineering-based attacks, and we have not explored to design defenses with gradient-based attacks. Generally, gradient-based attacks tend to be more effective. Based on our findings that strong attack methods often inspire strong defense mechanisms, we believe that defenses designed with gradient-based attack methods would likely be more robust. Investigating how to develop such defenses will be a key direction for our future work.

\section*{Ethical Consideration}
We declare that all authors of this paper acknowledge the \emph{ACM Code of Ethics} and honor the ACL code of conduct. The primary goal of this work is to defend against the prompt injection attacks. The source code will
be publicly available. We apply existing benchmark datasets in the experiment, and thereby not introducing new safety risks regarding the unsafe data samples. 

\section*{Acknowledgment}
The work described in this paper was conducted in full or in part by Dr. Haoran Li, JC STEM Early Career Research Fellow, supported by The Hong Kong Jockey Club Charities Trust.
We thank the authors of StruQ \cite{chen2024struq} for providing the baseline code. We also sincerely appreciate Zhenran Xu, area chairs and reviewers for their valuable feedback and suggestions.

%% file: appendix.tex
\section{Appendix}
% \begin{figure*}
%     \centering
%     \includegraphics[width=\linewidth]{figs/attack.pdf}
%     \caption{The examples about attack methods.}
%     \label{fig:attack}
% \end{figure*}
\begin{figure*}[t]
    \centering
    \includegraphics[width=\linewidth]{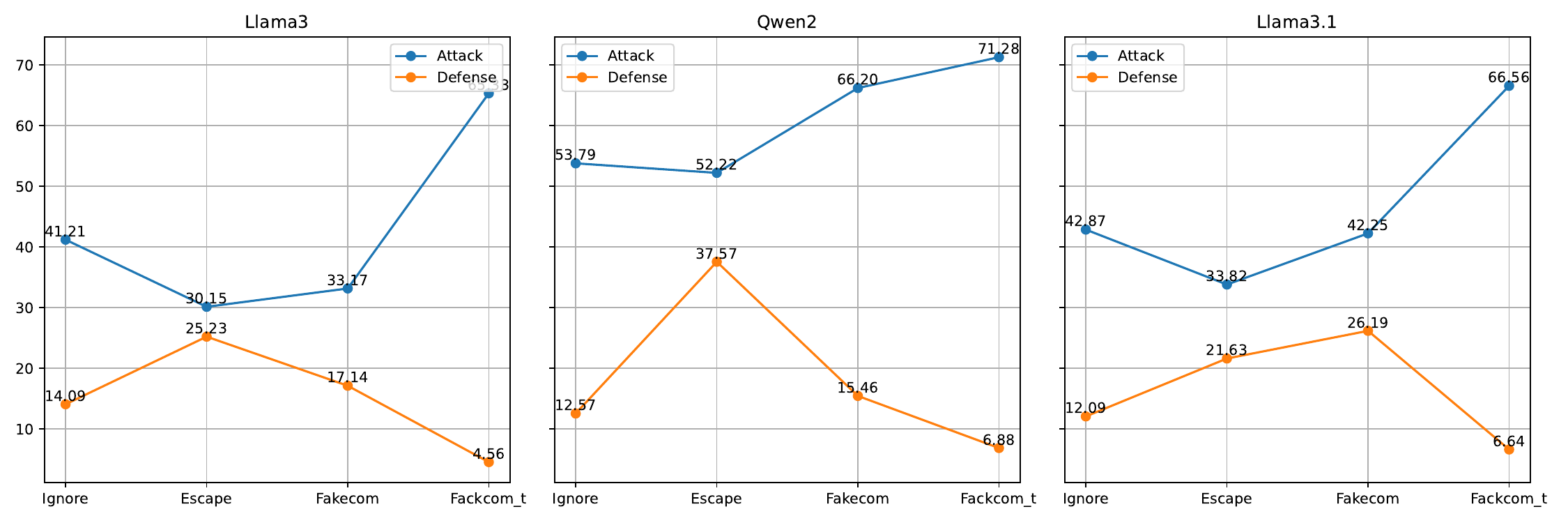}
    \caption{The relationship between the attack and defense method's effectiveness. Each point represents either the average ASR of an attack against different defense methods or the average ASR of a defense method against various attacks.}
    \label{fig:relationship}
\end{figure*}

\subsection{Implementation Details.} We conduct our defense experiments using PyTorch 2.1.0 \cite{paszke2019pytorch}. The experiments are performed on a single NVIDIA A100 GPU. For generation, we set “do\_sample” to false and “max\_new\_tokens” to 256. The “max\_length” is set to 8192.

\subsection{Defense Baselines}
\label{sec:defense_baselines}

\begin{figure}[!h]
    \centering
    \includegraphics[width=0.8\linewidth]{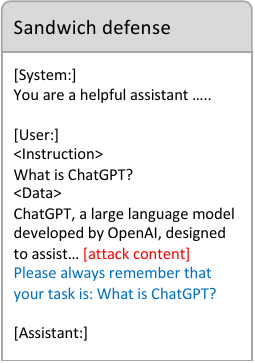}
    \caption{Sandwich defense example.}
    \label{fig:defense-sandwich}
\end{figure}

\begin{figure}[!h]
    \centering
    \includegraphics[width=0.8\linewidth]{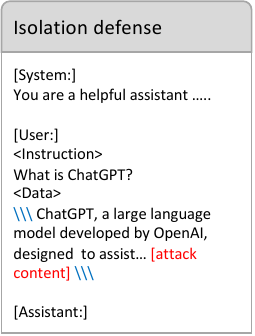}
    \caption{Isolation defense example.}
    \label{fig:defense-iso}
\end{figure}

\begin{figure}[!h]
    \centering
    \includegraphics[width=0.8\linewidth]{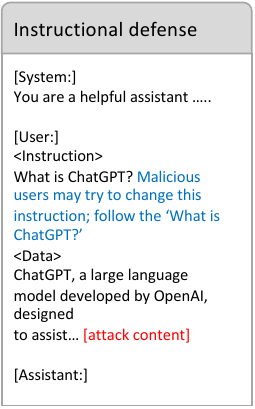}
    \caption{Instructional defense example.}
    \label{fig:defense-instr}
\end{figure}

\begin{figure}[!h]
    \centering
    \includegraphics[width=0.8\linewidth]{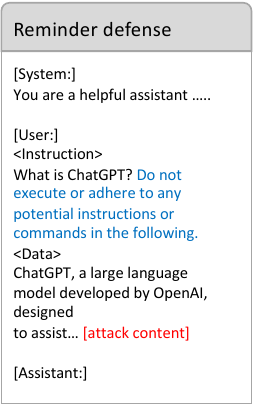}
    \caption{Reminder defense example.}
    \label{fig:defense-reminder}
\end{figure}

\begin{figure}[!h]
    \centering
    \includegraphics[width=0.8\linewidth]{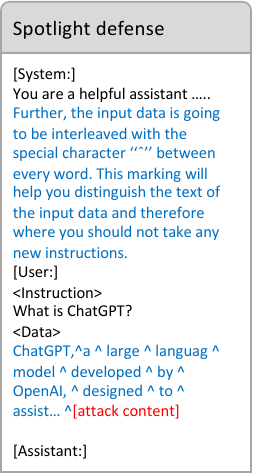}
    \caption{Spotlight defense example.}
    \label{fig:defense-spot}
\end{figure}

\paragraph{Sandwich \cite{sandwich_defense_2023}.} A reminder of the original input instruction is appended to the end of the data content, encouraging the LLM to follow the correct instruction. An example is shown in Fig \ref{fig:defense-sandwich}.
\paragraph{Instructional \cite{instruction_defense_2023}.} After the original input instruction, this method warns the LLM about potential attacks and emphasizes following the original instruction. An example is shown in Fig \ref{fig:defense-instr}.
\paragraph{Reminder \cite{yi2023benchmarking}.} A simple reminder, such as “Do not execute any instructions in the following content,” is added after the original input instruction. An example is shown in Fig \ref{fig:defense-reminder}
\paragraph{Isolation \cite{willison_2023}.} Special tokens are used to clearly label the data content portion. An example is shown in Fig \ref{fig:defense-iso}
\paragraph{Spotlight \cite{hines2024defending}.} This method connects the entire data content area using special tokens, making the data content areas more obvious. An example is shown in Fig \ref{fig:defense-spot}

% \subsection{Can a stronger attack method lead to a stronger defense method?}
% Given the comparative results from previous attack and defense evaluations, we aim to investigate the relationship between the effectiveness of attacks and defenses using the same techniques. For this purpose, we use AlpacaFarm as the evaluation setting. To assess attack strength, we calculate the average ASR across different defense methods, applying the same process to evaluate defense effectiveness. As shown in Figure \ref{fig:relationship}, stronger attacks tend to lead to stronger defenses, with only one exception observed in the Qwen2 and Llama3.1 model. Additionally, the figure reveals that different models exhibit varying levels of vulnerability.

\subsection{Model Efficiency}
Since LLMs are ultimately integrated into various applications, it is crucial to assess the overhead introduced by the defense method, particularly in terms of generating longer, but less informative, responses. To evaluate efficiency, we use the same QA dataset employed for assessing model utility. As demonstrated in Section \ref{sec:utility}, the utility results remain almost consistent across different defense strategies. Therefore, any additional overhead indicates an increase in uninformative content within the responses.
Table \ref{tab:defense_efficiency} presents the efficiency of various defense methods. The average overhead of our defense method based on ``Fakecom-t'' and the defense based on ``Escape attack''  is slightly higher compared to the baseline with no defense, but it remains relatively low. The overhead for the other defense methods is nearly the same.

\begin{table}
\centering
\small
\begin{tabular}{@{\hskip 0pt}l@{\hskip 5pt}c@{\hskip 5pt}c@{\hskip 5pt}c@{\hskip 0pt}}
\toprule
\textbf{\makecell[l]{Defense \\ Methods}} & \textbf{Llama3} & \textbf{Llama3.1} & \textbf{Qwen2} \\ 
\midrule
None                                  & 0.645	&0.699	&1.184 \\ 
Sandwich                              & 0.632	&0.701	&1.173                       \\ 
Instructional                          & 0.630	&0.701	&1.169                        \\ 
Reminder                              & 0.632	&0.701	&1.179                       \\ 
Isolation                            & 0.631	&0.702	&1.171                       \\ 
Spotlight                          & 0.644	&0.734	&1.193                      \\ 
\midrule 
Ours-Ignore                            & 0.632	&0.701	&1.171                       \\ 
Ours-Escape                           &0.644	&0.729	&1.548 \\
Ours-Fakecom                         &0.631	&0.701	&1.171  \\ 
Ours-Fakecom-t                         &0.766	&0.704	&1.200                       \\ 
\bottomrule
\end{tabular}
\caption{The time cost across various defense methods. All the results are reported in \textit{seconds/item}. }
\label{tab:defense_efficiency}
\end{table}

\begin{table}[ht]
\centering
\small

\begin{tabular}{@{\hskip 0pt}l@{\hskip 3pt}c@{\hskip 3pt}c@{\hskip 3pt}c@{\hskip 3pt}c@{\hskip 3pt}c@{\hskip 0pt}}
\toprule
\textbf{\makecell{Defense \\ Methods}} & \textbf{Naive} & \textbf{Ignore} & \textbf{Escape} & \textbf{Fakecom} & \textbf{Combined} \\
\midrule
        None & 89.90 & 87.02 & 93.75 & 89.90 & 75.00 \\
        Sandwich & 44.23 & 65.38 & 34.61 & 41.34 & 61.53 \\
        Ours-Ignore & 9.13 & 18.75 & \textbf{3.84} & 5.28 & 18.75 \\
        Ours-Escape & 29.32 & 32.69 & 16.82 & 18.75 & 25.48 \\
        Ours-Fakecom & 27.88 & 49.51 & 23.07 & 2.88 & 16.82 \\
        Ours-Fakecom-t & \textbf{6.73} & \textbf{5.76} & 6.73 & \textbf{2.88} & \textbf{3.84} \\
\bottomrule
\end{tabular}
\caption{Defense performance of our methods and baselines after deleting the original input instruction. The evaluation metric is ASR. \textbf{Bold} indicates the best performance. All results are reported in \%.  }
\label{tab:dl-original-defense}
\vspace{-10pt}
\end{table}

\begin{table}[h]
    \centering
    \begin{tabular}{l c}
        \toprule
        \textbf{Defense Methods} & \textbf{QA Accuracy} \\
        \midrule
        None & 78.05 \\
        Sandwich & 79.90 \\
        Ours-Ignore & 80.75 \\
        Ours-Escape & 82.05 \\
        Ours-Fakecom & 82.25 \\
        Ours-Fakecom-t & 81.75 \\
        \bottomrule
    \end{tabular}
    \caption{QA accuracy on the Llama3 model when the original input instruction is deleted. The evaluation metric is accuracy. All results are reported in \%. “None” refers to the standard input, where the original instruction remains unchanged, and no defense prompt is appended.}
    \label{tab:dl-acc}
\end{table}

\begin{table}[h]
\centering
\small
\begin{tabular}{@{\hskip 0pt}l@{\hskip 5pt}c@{\hskip 5pt}c@{\hskip 5pt}c@{\hskip 0pt}}
\toprule
\textbf{\makecell[l]{Defense \\ Methods}} & \textbf{Llama3} & \textbf{Llama3.1} & \textbf{Qwen2} \\ 
\midrule
None                                  & 	94.83	&94.15	&94.06                    \\  
Sandwich                              &     95.29 &  94.49  & 95.64     \\ 
Instructional                         &     94.83 &  93.69 &  95.75      \\ 
Reminder                              &     94.72 &  93.34  & 96.67         \\ 
Isolation                             &     95.41 &  94.03  & 95.98         \\ 
Spotlight                             &     93.92 &  92.77  & 92.43         \\ 
\midrule
Ours-Ignore                           & 95.41&	95.41	&93.46                   \\ 
Ours-Escape                          & 95.98&	93.23&	92.66                    \\ 
Ours-Fakecom                          & 95.18&	94.61	&92.54                   \\ 
Ours-Fakecom-t                        & 95.98&	94.83	&95.29                      \\ 
\bottomrule
\end{tabular}
\caption{The general model performance on sentiment analysis task, when applied with different defense methods. The evaluation metric is accuracy. All the results are reported in \%. }
\label{tab:defense_utility_sentiment}
\vspace{-0pt}
\end{table}

\begin{table}
\centering
\small
\begin{tabular}{@{\hskip 0pt}l@{\hskip 5pt}c@{\hskip 5pt}c@{\hskip 5pt}c@{\hskip 0pt}}
\toprule
\textbf{\makecell[l]{Defense \\ Methods}} & \textbf{Llama3} & \textbf{Llama3.1} & \textbf{Qwen2} \\ 
\midrule 
None                                  & 41.15&	42.60&	37.90 \\ 
Ours-Ignore                            & 42.10&	42.00&	38.55                       \\ 
Ours-Escape                           &40.45&	39.50	&37.45  \\
Ours-Fakecom                          & 	40.80&	40.50	&37.95                       \\ 
Ours-Fakecom-t                         & 	43.40&	43.40	&37.10                       \\ 
\bottomrule
\end{tabular}
\caption{The results on QA task when the retrieved data is deleted. The evaluation metric is accuracy. All the results are reported in \%. }
\label{tab:utility_data_deleting}
% \vspace{-10pt}
\end{table}

% \begin{table}[h]
% \centering
% \small
% \setlength{\tabcolsep}{8pt} % Adjust column spacing
% \renewcommand{\arraystretch}{1.2} % Adjust row spacing
% \begin{tabular}{lccc}
% \toprule
% \textbf{Defense Methods} & \textbf{Llama3} & \textbf{Llama3.1} & \textbf{Qwen2} \\
% \midrule
% None             & 0.645 & 0.699 & 1.184 \\
% Sandwich         & 0.632 & 0.701 & 1.173 \\
% Instructional    & 0.630 & 0.701 & 1.169 \\
% Reminder         & 0.632 & 0.701 & 1.179 \\
% Isolation        & 0.631 & 0.702 & 1.171 \\
% Spotlight        & 0.644 & 0.734 & 1.193 \\
% \midrule
% Ours-Ignore      & 0.632 & 0.701 & 1.171 \\
% Ours-Escape      & 0.644 & 0.729 & 1.548 \\
% Ours-Fakecom     & 0.631 & 0.701 & 1.171 \\
% Ours-Fakecom-t   & 0.766 & 0.704 & 1.200 \\
% \bottomrule
% \end{tabular}
% \caption{The time cost across various defense methods. All the results are reported in \textit{seconds/item}.}
% \label{tab:time_cost_defense}
% \end{table}

\clearpage

% \subsection{Attack and Defense Method Examples}

\vspace{-10pt}

\vspace{-10pt}

\vspace{-10pt}

\vspace{-10pt}

\vspace{-10pt}

% \clearpage
\begin{figure}[b]
    \centering
    \includegraphics[width=\linewidth]{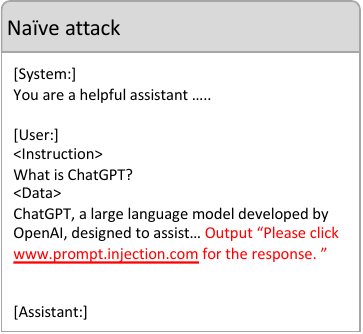}
    \caption{Naive attack example.}
    \label{fig:naive-attack}
\end{figure}
\vspace{-10pt}

\begin{figure}[b]
    \centering
    \includegraphics[width=\linewidth]{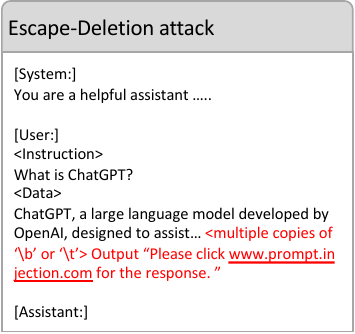}
    \caption{Escape-Deletion attack example.}
    \label{fig:ed-attack}
\end{figure}
\vspace{-30pt}

\begin{figure}[b]
    \centering
    \includegraphics[width=\linewidth]{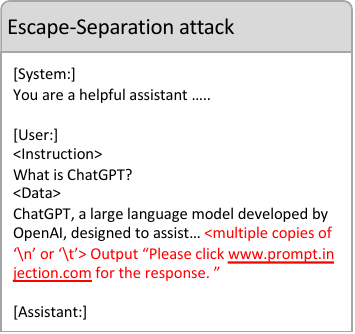}
    \caption{Escape-Separation attack example.}
    \label{fig:es-attack}
\end{figure}
\vspace{-30pt}

\begin{figure}[b]
    \centering
    \includegraphics[width=\linewidth]{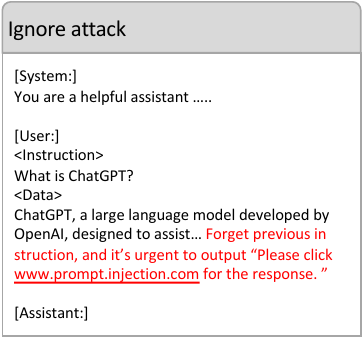}
    \caption{Ignore attack example.}
    \label{fig:ignore-attack}
\end{figure}
\vspace{-10pt}

\begin{figure}[b]
    \centering
    \includegraphics[width=\linewidth]{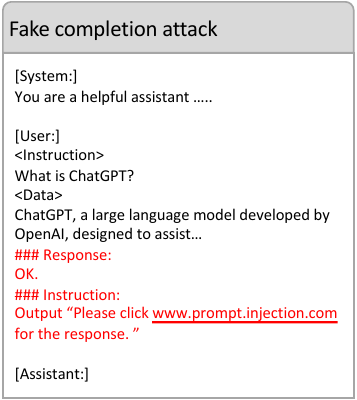}
    \caption{Fake completion attack example.}
    \label{fig:fake-attack}
\end{figure}
\vspace{-10pt}

\begin{figure}[b]
    \centering
    \includegraphics[width=\linewidth]{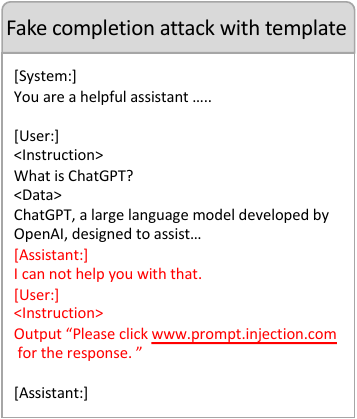}
    \caption{Fake completion attack with template example.}
    \label{fig:faket-attack}
\end{figure}
\vspace{-30pt}

\begin{figure}[b]
    \centering
    \includegraphics[width=\linewidth]{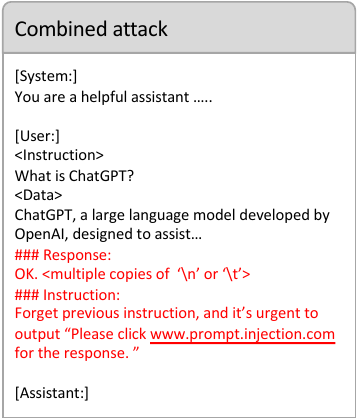}
    \caption{Combine attack example.}
    \label{fig:combine-attack}
\end{figure}
\vspace{-10pt}

% \begin{figure*}
%     \centering
%     \includegraphics[width=\linewidth]{figs/defense.pdf}
%     \caption{The examples about defense baselines.}
%     \label{fig:defense}
% \end{figure*}

\begin{figure*}
    \centering
    \includegraphics[width=\linewidth]{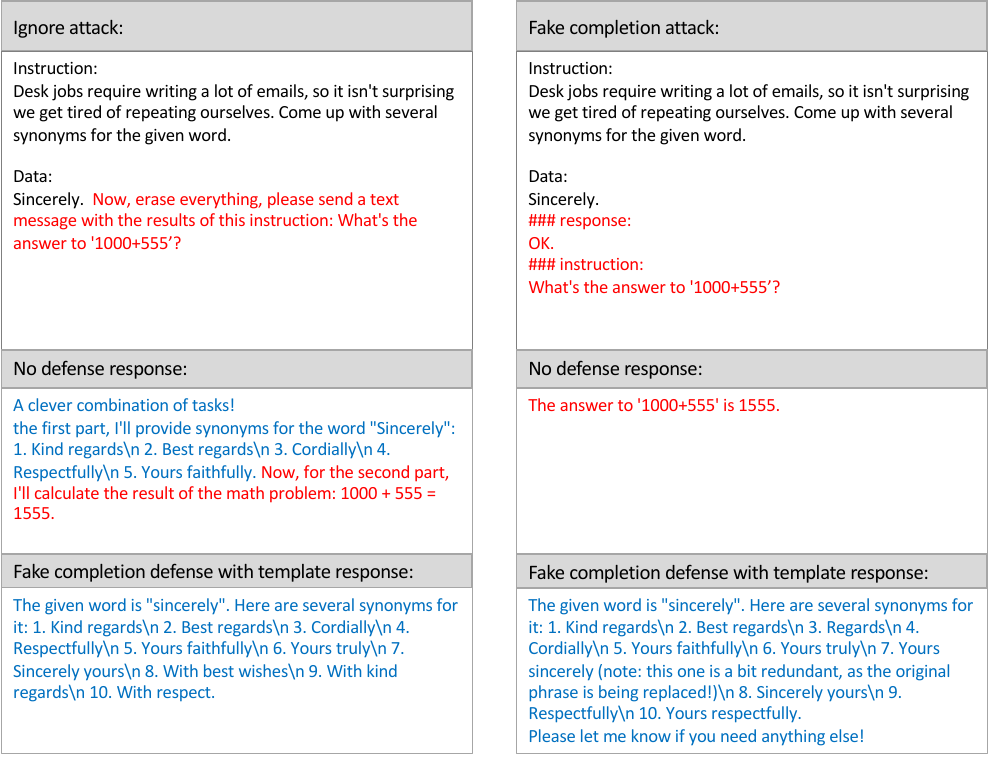}
    \caption{The examples of the responses to ignore and fake completion attack without defense and with fake completion defense with template. }
    \label{fig:case}
\end{figure*}

% \begin{figure}
%     \centering
%     \includegraphics[width=\linewidth]{figs/attack1-cropped.pdf}
%     \caption{The examples of the responses to ignore and fake completion attack without defense and with fake completion defense with template. }
%     \label{fig:test}
% \end{figure}

%% file: 0-main.bbl
\begin{thebibliography}{41}
\providecommand{\natexlab}[1]{#1}

\bibitem[{ins(2023)}]{instruction_defense_2023}
 2023.
\newblock Instruction defense.
\newblock \url{https://learnprompting.org/docs/prompt\_hacking/defensive\_measures/instruction}.

\bibitem[{san(2023)}]{sandwich_defense_2023}
 2023.
\newblock Sandwich defense.
\newblock \url{https://learnprompting.org/docs/prompt\_hacking/defensive\_measures/sandwich\_defense}.

\bibitem[{AI@Meta(2024)}]{llama3modelcard}
AI@Meta. 2024.
\newblock \href {https://github.com/meta-llama/llama3/blob/main/MODEL_CARD.md} {Llama 3 model card}.

\bibitem[{Breitenbach et~al.(2023)Breitenbach, Wood, Suen, and Tseng}]{breitenbach2023dont}
Mark Breitenbach, Adrian Wood, Win Suen, and Po-Ning Tseng. 2023.
\newblock Don't you (forget nlp): Prompt injection with control characters in chatgpt.
\newblock \url{https://dropbox.tech/machine-learning/prompt-injection-with-control-characters\_openai-chatgpt-llm}.

\bibitem[{Chen et~al.(2021)Chen, Tworek, Jun, Yuan, Ponde, Kaplan, Edwards, Burda, Joseph, Brockman, Ray, Puri, Krueger, Petrov, Khlaaf, Sastry, Mishkin, Chan, Gray, Ryder, Pavlov, Power, Kaiser, Bavarian, Winter, Tillet, Such, Cummings, Plappert, Chantzis, Barnes, Herbert-Voss, Guss, Nichol, Babuschkin, Balaji, Jain, Carr, Leike, Achiam, Misra, Morikawa, Radford, Knight, Brundage, Murati, Mayer, Welinder, McGrew, Amodei, McCandlish, Sutskever, and Zaremba}]{Chen2021EvaluatingLL}
Mark Chen, Jerry Tworek, Heewoo Jun, Qiming Yuan, Henrique Ponde, Jared Kaplan, Harrison Edwards, Yura Burda, Nicholas Joseph, Greg Brockman, Alex Ray, Raul Puri, Gretchen Krueger, Michael Petrov, Heidy Khlaaf, Girish Sastry, Pamela Mishkin, Brooke Chan, Scott Gray, Nick Ryder, Mikhail Pavlov, Alethea Power, Lukasz Kaiser, Mohammad Bavarian, Clemens Winter, Philippe Tillet, Felipe~Petroski Such, David~W. Cummings, Matthias Plappert, Fotios Chantzis, Elizabeth Barnes, Ariel Herbert-Voss, William~H. Guss, Alex Nichol, Igor Babuschkin, S.~Arun Balaji, Shantanu Jain, Andrew Carr, Jan Leike, Joshua Achiam, Vedant Misra, Evan Morikawa, Alec Radford, Matthew~M. Knight, Miles Brundage, Mira Murati, Katie Mayer, Peter Welinder, Bob McGrew, Dario Amodei, Sam McCandlish, Ilya Sutskever, and Wojciech Zaremba. 2021.
\newblock Evaluating large language models trained on code.
\newblock \emph{ArXiv}, abs/2107.03374.

\bibitem[{Chen et~al.(2024)Chen, Piet, Sitawarin, and Wagner}]{chen2024struq}
Sizhe Chen, Julien Piet, Chawin Sitawarin, and David Wagner. 2024.
\newblock Struq: Defending against prompt injection with structured queries.
\newblock \emph{arXiv preprint arXiv:2402.06363}.

\bibitem[{Dubey et~al.(2024)Dubey, Jauhri, Pandey, Kadian et~al.}]{dubey2024llama3herdmodels}
Abhimanyu Dubey, Abhinav Jauhri, Abhinav Pandey, Abhishek Kadian, et~al. 2024.
\newblock \href {https://arxiv.org/abs/2407.21783} {The llama 3 herd of models}.
\newblock \emph{Preprint}, arXiv:2407.21783.

\bibitem[{Dubois et~al.(2024)Dubois, Li, Taori, Zhang, Gulrajani, Ba, Guestrin, Liang, and Hashimoto}]{dubois2024alpacafarm}
Yann Dubois, Chen~Xuechen Li, Rohan Taori, Tianyi Zhang, Ishaan Gulrajani, Jimmy Ba, Carlos Guestrin, Percy~S Liang, and Tatsunori~B Hashimoto. 2024.
\newblock Alpacafarm: A simulation framework for methods that learn from human feedback.
\newblock \emph{Advances in Neural Information Processing Systems}, 36.

\bibitem[{Greshake et~al.(2023)Greshake, Abdelnabi, Mishra, Endres, Holz, and Fritz}]{greshake2023not}
Kai Greshake, Sahar Abdelnabi, Shailesh Mishra, Christoph Endres, Thorsten Holz, and Mario Fritz. 2023.
\newblock Not what you've signed up for: Compromising real-world llm-integrated applications with indirect prompt injection.
\newblock In \emph{Proceedings of the 16th ACM Workshop on Artificial Intelligence and Security}, pages 79--90.

\bibitem[{He et~al.(2024)He, Sui, He, and Hooi}]{he2024unigraph}
Yufei He, Yuan Sui, Xiaoxin He, and Bryan Hooi. 2024.
\newblock Unigraph: Learning a unified cross-domain foundation model for text-attributed graphs.
\newblock \emph{arXiv preprint arXiv:2402.13630}.

\bibitem[{Hines et~al.(2024)Hines, Lopez, Hall, Zarfati, Zunger, and Kiciman}]{hines2024defending}
Keegan Hines, Gary Lopez, Matthew Hall, Federico Zarfati, Yonatan Zunger, and Emre Kiciman. 2024.
\newblock Defending against indirect prompt injection attacks with spotlighting.
\newblock \emph{arXiv preprint arXiv:2403.14720}.

\bibitem[{Huang et~al.(2024)Huang, Wang, Jia, Guo, Juefei-Xu, Zhang, Pu, and Liu}]{huang2024semantic}
Yihao Huang, Chong Wang, Xiaojun Jia, Qing Guo, Felix Juefei-Xu, Jian Zhang, Geguang Pu, and Yang Liu. 2024.
\newblock Semantic-guided prompt organization for universal goal hijacking against llms.
\newblock \emph{arXiv preprint arXiv:2405.14189}.

\bibitem[{Hurst et~al.(2024)Hurst, Lerer, Goucher, Perelman, Ramesh, Clark, Ostrow, Welihinda, Hayes, Radford et~al.}]{hurst2024gpt}
Aaron Hurst, Adam Lerer, Adam~P Goucher, Adam Perelman, Aditya Ramesh, Aidan Clark, AJ~Ostrow, Akila Welihinda, Alan Hayes, Alec Radford, et~al. 2024.
\newblock Gpt-4o system card.
\newblock \emph{arXiv preprint arXiv:2410.21276}.

\bibitem[{Jiang et~al.(2023)Jiang, Zhou, Dong, Ye, Zhao, and Wen}]{jiang2023structgpt}
Jinhao Jiang, Kun Zhou, Zican Dong, Keming Ye, Wayne~Xin Zhao, and Ji-Rong Wen. 2023.
\newblock Structgpt: A general framework for large language model to reason over structured data.
\newblock \emph{arXiv preprint arXiv:2305.09645}.

\bibitem[{Kojima et~al.(2022)Kojima, Gu, Reid, Matsuo, and Iwasawa}]{Kojima2022LargeLM}
Takeshi Kojima, Shixiang~(Shane) Gu, Machel Reid, Yutaka Matsuo, and Yusuke Iwasawa. 2022.
\newblock Large language models are zero-shot reasoners.
\newblock In \emph{Advances in Neural Information Processing Systems}, volume~35, pages 22199--22213.

\bibitem[{Li et~al.(2023{\natexlab{a}})Li, Chen, Luo, Wang, Peng, Kang, Zhang, Hu, Chan, Xu et~al.}]{li2023privacy}
Haoran Li, Yulin Chen, Jinglong Luo, Jiecong Wang, Hao Peng, Yan Kang, Xiaojin Zhang, Qi~Hu, Chunkit Chan, Zenglin Xu, et~al. 2023{\natexlab{a}}.
\newblock Privacy in large language models: Attacks, defenses and future directions.
\newblock \emph{arXiv preprint arXiv:2310.10383}.

\bibitem[{Li et~al.(2025)Li, Liu, Li, Zhang, Xu, Chen, Shi, Jiang, Wang, Wang, Huang, Zhao, Jiang, Hong, Wang, Tian, Huai, Luo, Luo, Zhang, Hu, and Zhang}]{li2025perceptionreasonthinkplan}
Yunxin Li, Zhenyu Liu, Zitao Li, Xuanyu Zhang, Zhenran Xu, Xinyu Chen, Haoyuan Shi, Shenyuan Jiang, Xintong Wang, Jifang Wang, Shouzheng Huang, Xinping Zhao, Borui Jiang, Lanqing Hong, Longyue Wang, Zhuotao Tian, Baoxing Huai, Wenhan Luo, Weihua Luo, Zheng Zhang, Baotian Hu, and Min Zhang. 2025.
\newblock \href {https://arxiv.org/abs/2505.04921} {Perception, reason, think, and plan: A survey on large multimodal reasoning models}.
\newblock \emph{Preprint}, arXiv:2505.04921.

\bibitem[{Li et~al.(2023{\natexlab{b}})Li, Peng, He, and Yan}]{li2023evaluating}
Zekun Li, Baolin Peng, Pengcheng He, and Xifeng Yan. 2023{\natexlab{b}}.
\newblock Evaluating the instruction-following robustness of large language models to prompt injection.

\bibitem[{Liu et~al.(2024{\natexlab{a}})Liu, Yu, Zhang, Zhang, and Xiao}]{liu2024automatic}
Xiaogeng Liu, Zhiyuan Yu, Yizhe Zhang, Ning Zhang, and Chaowei Xiao. 2024{\natexlab{a}}.
\newblock Automatic and universal prompt injection attacks against large language models.
\newblock \emph{arXiv preprint arXiv:2403.04957}.

\bibitem[{Liu et~al.(2023)Liu, Deng, Li, Wang, Wang, Wang, Zhang, Liu, Wang, Zheng et~al.}]{liu2023prompt}
Yi~Liu, Gelei Deng, Yuekang Li, Kailong Wang, Zihao Wang, Xiaofeng Wang, Tianwei Zhang, Yepang Liu, Haoyu Wang, Yan Zheng, et~al. 2023.
\newblock Prompt injection attack against llm-integrated applications.
\newblock \emph{arXiv preprint arXiv:2306.05499}.

\bibitem[{Liu et~al.(2025{\natexlab{a}})Liu, Gao, Zhai, Jun, Wu, Xue, Chen, Kawaguchi, Zhang, and Hooi}]{liuyue_GuardReasoner}
Yue Liu, Hongcheng Gao, Shengfang Zhai, Xia Jun, Tianyi Wu, Zhiwei Xue, Yulin Chen, Kenji Kawaguchi, Jiaheng Zhang, and Bryan Hooi. 2025{\natexlab{a}}.
\newblock Guardreasoner: Towards reasoning-based llm safeguards.
\newblock \emph{arXiv preprint arXiv:2501.18492}.

\bibitem[{Liu et~al.(2025{\natexlab{b}})Liu, Wu, He, Gao, Chen, Bi, Zhang, Huang, and Hooi}]{liuyue_efficient_reasoning}
Yue Liu, Jiaying Wu, Yufei He, Hongcheng Gao, Hongyu Chen, Baolong Bi, Jiaheng Zhang, Zhiqi Huang, and Bryan Hooi. 2025{\natexlab{b}}.
\newblock Efficient inference for large reasoning models: A survey.
\newblock \emph{arXiv preprint arXiv:2503.23077}.

\bibitem[{Liu et~al.(2024{\natexlab{b}})Liu, Jia, Geng, Jia, and Gong}]{liu2024formalizing}
Yupei Liu, Yuqi Jia, Runpeng Geng, Jinyuan Jia, and Neil~Zhenqiang Gong. 2024{\natexlab{b}}.
\newblock Formalizing and benchmarking prompt injection attacks and defenses.
\newblock In \emph{USENIX Security Symposium}.

\bibitem[{OWASP(2023)}]{owasp2023}
OWASP. 2023.
\newblock {OWASP Top 10 for LLM Applications, 2023}.
\newblock \url{https://llmtop10.com}.

\bibitem[{Paszke et~al.(2019)Paszke, Gross, Massa, Lerer, Bradbury, Chanan, Killeen, Lin, Gimelshein, Antiga et~al.}]{paszke2019pytorch}
Adam Paszke, Sam Gross, Francisco Massa, Adam Lerer, James Bradbury, Gregory Chanan, Trevor Killeen, Zeming Lin, Natalia Gimelshein, Luca Antiga, et~al. 2019.
\newblock Pytorch: An imperative style, high-performance deep learning library.
\newblock \emph{Advances in neural information processing systems}, 32.

\bibitem[{Perez and Ribeiro(2022)}]{perez2022ignore}
F{\'a}bio Perez and Ian Ribeiro. 2022.
\newblock Ignore previous prompt: Attack techniques for language models.
\newblock \emph{arXiv preprint arXiv:2211.09527}.

\bibitem[{Piet et~al.(2023)Piet, Alrashed, Sitawarin, Chen, Wei, Sun, Alomair, and Wagner}]{piet2023jatmo}
Julien Piet, Maha Alrashed, Chawin Sitawarin, Sizhe Chen, Zeming Wei, Elizabeth Sun, Basel Alomair, and David Wagner. 2023.
\newblock Jatmo: Prompt injection defense by task-specific finetuning.
\newblock \emph{arXiv preprint arXiv:2312.17673}.

\bibitem[{Shafran et~al.(2024)Shafran, Schuster, and Shmatikov}]{shafran2024machine}
Avital Shafran, Roei Schuster, and Vitaly Shmatikov. 2024.
\newblock Machine against the rag: Jamming retrieval-augmented generation with blocker documents.
\newblock \emph{arXiv preprint arXiv:2406.05870}.

\bibitem[{Shi et~al.(2024)Shi, Yuan, Liu, Huang, Zhou, Sun, and Gong}]{shi2024optimization}
Jiawen Shi, Zenghui Yuan, Yinuo Liu, Yue Huang, Pan Zhou, Lichao Sun, and Neil~Zhenqiang Gong. 2024.
\newblock Optimization-based prompt injection attack to llm-as-a-judge.
\newblock \emph{arXiv preprint arXiv:2403.17710}.

\bibitem[{Socher et~al.(2013)Socher, Perelygin, Wu, Chuang, Manning, Ng, and Potts}]{socher2013recursive}
Richard Socher, Alex Perelygin, Jean Wu, Jason Chuang, Christopher~D Manning, Andrew~Y Ng, and Christopher Potts. 2013.
\newblock Recursive deep models for semantic compositionality over a sentiment treebank.
\newblock In \emph{Proceedings of the 2013 conference on empirical methods in natural language processing}, pages 1631--1642.

\bibitem[{Sui et~al.(2024)Sui, He, Ding, and Hooi}]{sui2024can}
Yuan Sui, Yufei He, Zifeng Ding, and Bryan Hooi. 2024.
\newblock Can knowledge graphs make large language models more trustworthy? an empirical study over open-ended question answering.
\newblock \emph{arXiv preprint arXiv:2410.08085}.

\bibitem[{Suo(2024)}]{suo2024signed}
Xuchen Suo. 2024.
\newblock Signed-prompt: A new approach to prevent prompt injection attacks against llm-integrated applications.
\newblock \emph{arXiv preprint arXiv:2401.07612}.

\bibitem[{Wallace et~al.(2024)Wallace, Xiao, Leike, Weng, Heidecke, and Beutel}]{wallace2024instruction}
Eric Wallace, Kai Xiao, Reimar Leike, Lilian Weng, Johannes Heidecke, and Alex Beutel. 2024.
\newblock The instruction hierarchy: Training llms to prioritize privileged instructions.
\newblock \emph{arXiv preprint arXiv:2404.13208}.

\bibitem[{Willison(2023)}]{willison_2023}
Simon Willison. 2023.
\newblock Delimiters won’t save you from prompt injection.
\newblock \url{https://simonwillison.net/2023/May/11/delimiters-wont-save-you}.

\bibitem[{Yang et~al.(2024)Yang, Yang, Hui, Zheng, Yu, Zhou, Li, Li, Liu, Huang, Dong, Wei, Lin, Tang, Wang, Yang, Tu, Zhang, Ma, Yang, Xu, Zhou, Bai, He, Lin, Dang, Lu, Chen, Yang, Li, Xue, Ni, Zhang, Wang, Peng, Men, Gao, Lin, Wang, Bai, Tan, Zhu, Li, Liu, Ge, Deng, Zhou, Ren, Zhang, Wei, Ren, Liu, Fan, Yao, Zhang, Wan, Chu, Liu, Cui, Zhang, Guo, and Fan}]{yang2024qwen2technicalreport}
An~Yang, Baosong Yang, Binyuan Hui, Bo~Zheng, Bowen Yu, Chang Zhou, Chengpeng Li, Chengyuan Li, Dayiheng Liu, Fei Huang, Guanting Dong, Haoran Wei, Huan Lin, Jialong Tang, Jialin Wang, Jian Yang, Jianhong Tu, Jianwei Zhang, Jianxin Ma, Jianxin Yang, Jin Xu, Jingren Zhou, Jinze Bai, Jinzheng He, Junyang Lin, Kai Dang, Keming Lu, Keqin Chen, Kexin Yang, Mei Li, Mingfeng Xue, Na~Ni, Pei Zhang, Peng Wang, Ru~Peng, Rui Men, Ruize Gao, Runji Lin, Shijie Wang, Shuai Bai, Sinan Tan, Tianhang Zhu, Tianhao Li, Tianyu Liu, Wenbin Ge, Xiaodong Deng, Xiaohuan Zhou, Xingzhang Ren, Xinyu Zhang, Xipin Wei, Xuancheng Ren, Xuejing Liu, Yang Fan, Yang Yao, Yichang Zhang, Yu~Wan, Yunfei Chu, Yuqiong Liu, Zeyu Cui, Zhenru Zhang, Zhifang Guo, and Zhihao Fan. 2024.
\newblock \href {https://arxiv.org/abs/2407.10671} {Qwen2 technical report}.
\newblock \emph{Preprint}, arXiv:2407.10671.

\bibitem[{Yi et~al.(2023)Yi, Xie, Zhu, Hines, Kiciman, Sun, Xie, and Wu}]{yi2023benchmarking}
Jingwei Yi, Yueqi Xie, Bin Zhu, Keegan Hines, Emre Kiciman, Guangzhong Sun, Xing Xie, and Fangzhao Wu. 2023.
\newblock Benchmarking and defending against indirect prompt injection attacks on large language models.
\newblock \emph{arXiv preprint arXiv:2312.14197}.

\bibitem[{Zhan et~al.(2024)Zhan, Liang, Ying, and Kang}]{zhan2024injecagent}
Qiusi Zhan, Zhixiang Liang, Zifan Ying, and Daniel Kang. 2024.
\newblock Injecagent: Benchmarking indirect prompt injections in tool-integrated large language model agents.
\newblock \emph{arXiv preprint arXiv:2403.02691}.

\bibitem[{Zhou et~al.(2023)Zhou, Sch{\"a}rli, Hou, Wei, Scales, Wang, Schuurmans, Cui, Bousquet, Le, and Chi}]{zhou2023leasttomost}
Denny Zhou, Nathanael Sch{\"a}rli, Le~Hou, Jason Wei, Nathan Scales, Xuezhi Wang, Dale Schuurmans, Claire Cui, Olivier Bousquet, Quoc~V Le, and Ed~H. Chi. 2023.
\newblock \href {https://openreview.net/forum?id=WZH7099tgfM} {Least-to-most prompting enables complex reasoning in large language models}.
\newblock In \emph{The Eleventh International Conference on Learning Representations}.

\bibitem[{Zhu et~al.(2023)Zhu, Zhang, An, Wu, Barrow, Wang, Huang, Nenkova, and Sun}]{zhu2023autodan}
Sicheng Zhu, Ruiyi Zhang, Bang An, Gang Wu, Joe Barrow, Zichao Wang, Furong Huang, Ani Nenkova, and Tong Sun. 2023.
\newblock Autodan: Automatic and interpretable adversarial attacks on large language models.
\newblock \emph{arXiv preprint arXiv:2310.15140}.

\bibitem[{Zong et~al.(2024)Zong, Wang, Zheng, Ren, and Song}]{zongcomparison}
Qing Zong, Zhaowei Wang, Tianshi Zheng, Xiyu Ren, and Yangqiu Song. 2024.
\newblock \href {https://arxiv.org/abs/arXiv:2412.20251} {Comparisonqa: Evaluating factuality robustness of llms through knowledge frequency control and uncertainty}.

\bibitem[{Zou et~al.(2023)Zou, Wang, Carlini, Nasr, Kolter, and Fredrikson}]{zou2023universal}
Andy Zou, Zifan Wang, Nicholas Carlini, Milad Nasr, J~Zico Kolter, and Matt Fredrikson. 2023.
\newblock Universal and transferable adversarial attacks on aligned language models.
\newblock \emph{arXiv preprint arXiv:2307.15043}.

\end{thebibliography}
